# How many bits may fit in a single magnetic dot? XMCD-PEEM evidences the switching of Néel caps inside Bloch domain walls.

F. Cheynis[1], A. Masseboeuf[2], O. Fruchart[1], N. Rougemaille[1], J. C. Toussaint[1], R. Belkhou[3,4], P. Bayle-Guillemaud[2], A. Marty[5]

[1]*Institut NÉEL, CNRS et Université Joseph Fourier, Grenoble, France*

[2]*CEA-Grenoble, INAC/SP2M/LEMMA, Grenoble, France*

[3]*Synchrotron SOLEIL, L'Orme des Merisiers Saint-Aubin, Gif-sur-Yvette Cedex, France*

[4]*ELETTRA, Sincrotrone Trieste, Basovizza, Trieste, Italy*

[5]*CEA-Grenoble, INAC/SP2M/NM, 17 rue des Martyrs, Grenoble, France*

Email: Olivier.Fruchart@grenoble.cnrs.fr

Data storage relies on the handling of two states, called bits. The market of mass storage is currently still dominated by magnetic technology, hard disk drives for the broad public and tapes for massive archiving. In these devices each bit is stored in the form of the direction of magnetization of nanosized magnetic domains, i.e. areas of ferromagnetic materials displaying a uniform magnetization. While miniaturization is the conventional way to fuel the continuous increase of device density, disruptive solutions are also sought. To these pertain in recent years many fundamental studies no longer considering the magnetic domains themselves, but the manipulation of the domain walls (DWs) that separate such domains. Concepts of storage and logic based on the propagation of DWs along lithographically-patterned stripes have been patented, while many fundamental aspects of DW propagation deeply related to condensed matter physics are still hotly debated. If one now considers magnetic dots of submicrometer dimensions, the magnetization has a tendency to curl along the outer edges of the nanostructure to close its magnetic flux and thereby reduce its magnetostatic energy. Then both domains and DWs of well-defined geometries arise, whose combined manipulation has been proposed as a multilevel magnetic storage scheme. For example in circular dots the chirality of the flux closure and the vertical polarity of the central magnetic vortex (a domain wall reduced to a tube instead of a surface) are two independent bits, whose manipulation has led to numerous studies in the past few years.

In our study we have been one step further and demonstrated the manipulation of dots with three degrees of freedom. This was achieved considering elongated dots, in which case the central vortex is replaced by an elongated DW. Micrometer-sized faceted epitaxial Fe(110) dots self-assembled under ultra-high vacuum and prepared using pulsed laser deposition have been used as model systems [1] (FIG1a). Beyond its vertical polarity which is analogous to that of a magnetic vortex, such a domain wall possesses an extra degree of freedom: at both upper and lower dot surfaces the magnetization of the DW turn in-plane to minimize magnetostatic energy. These areas are called Néel caps (NCs) and are antiparallel to each other in the absence of external field, again to minimize magnetostatic energy (FIG.2a-c) [2]. The extra degree of freedom is therefore whether these are in a (-,+) or (+,-) state, named after the direction of bottom and top NCs (FIG2d). The potential effect of external fields on NCs had been discussed decades ago in the case of thin films, however never unambiguously demonstrated. We predicted by simulation that NCs should turn parallel and aligned upon application of an external field along the short axis of the dots, while both chirality and DW polarity should remain unchanged [states (+,+) or (-,-) depending on the sign of the applied field]. Then upon decreasing the field back to zero state (+,+) switches back to (+,-), while state (-,-) switches back to (-,+) (FIG2e). This selection process is made possible in a dot with tilted facets, not in thin films, explaining the above-mentioned long-standing unfruitful experimental search in thin films.

Our experimental confirmation came from the use of the French-Soleil LEEM-PEEM Elmitech instrument currently hosted at the Nanospectroscopy beamline. Based on a LEEM instrument [Low-Energy Electron Microscope, (FIG.2b)] and exploiting the magnetic sensitivity owing to X-ray Magnetic Circular Dichroism (XMCD), the collection of emitted photoelectrons (PEEM) allows one to build surface maps of a magnetization component with a spatial resolution around 25nm. NCs are then revealed as a thin stripe of dark or light contrast along the length of the dots, depending on the (+,-) or (-,+) state (FIG.2c). The use of XMCD-PEEM proved to be crucial as the usual technique of Magnetic Force Microscopy (MFM) is essentially sensitive to the polarity of the DW, and does not probe NCs. As magnetic fields are hardly compatible with this low-energy electrons the magnetization process was performed off-stage, and observed a posteriori. For each field several tens of dots were imaged (FIG1b-c). This way the switching process could be observed and the mean switching field of 100mT was found, in close agreement with the simulations [3]. Closely later the process could be monitored in real time under applied field using Lorentz Microscopy [3], however again the seminal XMCD-PEEM investigation was crucial as in Lorentz Microscopy only indirect information about the switching field is gained, NCs are not seen, like for MFM.

Beyond potential futuristic applications in data storage, this study should mainly excite our physicist curiosity about the extension of the concept of magnetization reversal to the inner structure of DWs, beyond the classical case of extended domains, and trigger new studies in this direction.

Figures:

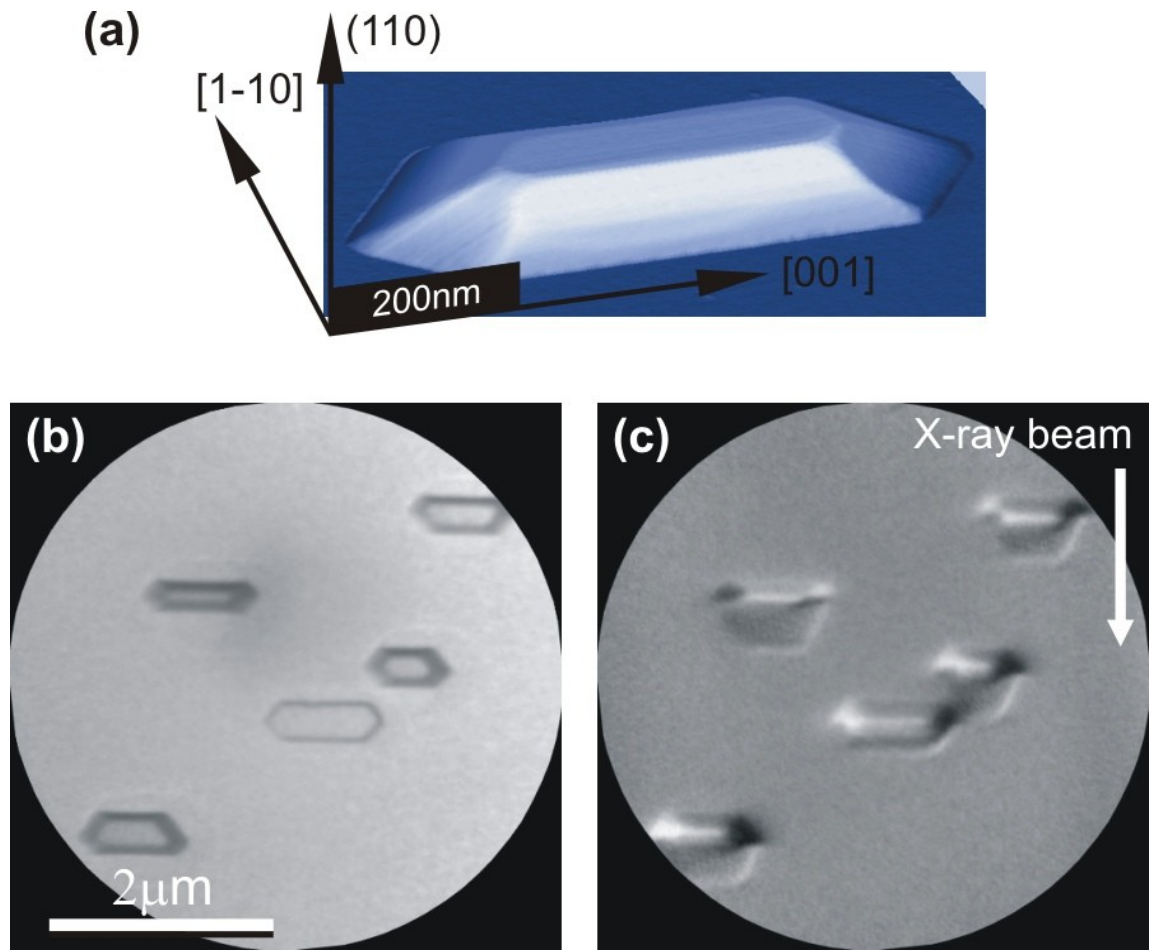

**Figure 1:** (a) Three-dimensional view of a typical self-assembled epitaxial Fe(110) dot (atomic force microscopy, true aspect ratios). (b) LEEM (topography) and (c) XMCD-PEEM (magnetism) typical view of an ensemble of dots. After magnetization at -130mT the dots are in the (-,+) state at remanence whatever their size, height, or aspect ratio. The white arrow indicates the direction of the x rays, thus the component of in-plane surface magnetization imaged.

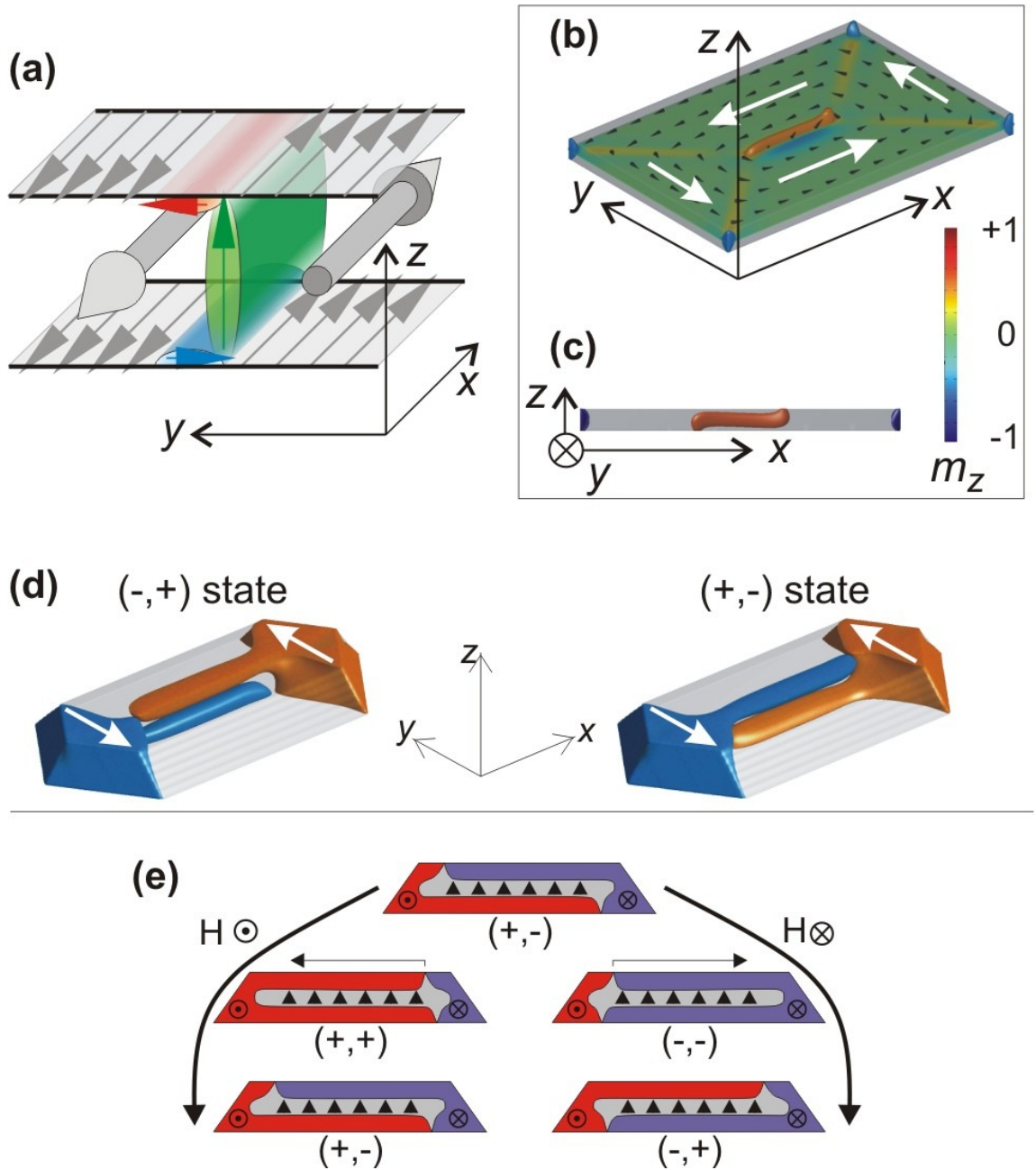

**Figure 2:** (a) Schematic view of a domain wall of Bloch type, terminated by a Néel cap at each surface. The magnetization is opposite in the top and bottom Néel caps, and points along the ±y axes, i.e., across the plane of the wall. (b–c) 3D and cross section views of a flux-closure state in a rectangular magnetic dot (700x500x50 nm). In (b) only volumes with normalized perpendicular magnetization $|m_z| > 0.5$ are displayed, which highlights the domain wall. The in-plane curling of magnetization is indicated by white arrows. (d) Simulation in an elongated dot of the (-,+) and (+,-) states. Only volumes with $|m_y| > 0.5$ are displayed, while positive and negative values appear red and blue, respectively. (e) Schematic cross-sectional view of the switching of NCs: the final state is (-,+) or (+,-) depending on the sign of the applied field.